\documentclass[aps,prl,floatfix, superscriptaddress, nofootinbib]{revtex4} 

\usepackage{graphicx}
\usepackage{graphics}
\usepackage{amssymb}
\usepackage{amsmath}
\usepackage{bm}
\usepackage{epsfig}
\usepackage{latexsym}
\usepackage{color}
\usepackage{subfigure}

\newcommand{\ket}[1]{| #1\rangle}
\newcommand{\bra}[1]{\langle #1 |}

\newcommand{\tr}{{\rm tr \thinspace}}
\def\ketc[#1]{\vert #1 \rangle}
\def\brac[#1]{\langle #1 \vert}

\newcommand{\beq}{\begin{equation}}
\newcommand{\eeq}{\end{equation}}
\newcommand{\bqa}{\begin{eqnarray}}
\newcommand{\eqa}{\end{eqnarray}}
\newcommand{\nn}{\nonumber}

\newcommand{\erf}[1]{Eq.~(\ref{#1})}
\newcommand{\dg}{^\dagger}

\newcommand{\BQIC}{Berkeley Center for Quantum Information and Computation, Berkeley, California 94720 USA}
\newcommand{\DeptChem}{Department of Chemistry, University of California, Berkeley, California 94720 USA}
\newcommand{\PBDLBL}{Physical Bioscience Division, Lawrence Berkeley National Laboratory, Berkeley, California 94720 USA}

\begin{document}

\title{Quantum entanglement in photosynthetic light harvesting complexes}

\author{Mohan Sarovar}
\email{msarovar@berkeley.edu}
\affiliation{\BQIC} 	\affiliation{\DeptChem} 

\author{Akihito Ishizaki}
\affiliation{\DeptChem}	\affiliation{\PBDLBL}
 
\author{Graham R. Fleming}
\affiliation{\DeptChem}	\affiliation{\PBDLBL} 

\author{K. Birgitta Whaley}
\affiliation{\BQIC}		\affiliation{\DeptChem}

\begin{abstract}
Light harvesting components of photosynthetic organisms are complex, coupled, many-body quantum systems, in which electronic coherence has recently been shown to survive for relatively long time scales despite the decohering effects of their environments. Within this context, we analyze entanglement in multi-chromophoric light harvesting complexes, and establish methods for quantification of entanglement by presenting necessary and sufficient conditions for entanglement and by deriving a measure of global entanglement. These methods are then applied to the Fenna-Matthews-Olson (FMO) protein to extract the initial state and temperature dependencies of entanglement. We show that while FMO in 
natural conditions largely contains bipartite entanglement between dimerized chromophores, a small amount of long-range and multipartite entanglement exists even at physiological temperatures. This constitutes the first rigorous quantification of entanglement in a biological system. Finally, we discuss the practical utilization of entanglement in densely packed molecular aggregates such as light harvesting complexes.
\end{abstract}

\maketitle

Unlike in classical physics, within quantum mechanics one can have maximal knowledge of a composite physical system and still not be able to assign a definite state to its constituent elements without reference to their relation to each other \cite{Ein.Pod.etal-1935, Sch-1935}. Such systems are called entangled, and entanglement is a characteristic quantum mechanical effect that has been widely investigated in recent years \cite{Hor.Hor.etal-2009, Ami.Faz.etal-2008}. Entanglement is often viewed as a fragile and exotic property, and in the quantum information context, where it is used as a resource for information processing tasks, precisely engineered entangled states of interest can indeed be both fragile and difficult to manufacture. However, it has also been recognized that entanglement is a natural feature of coherent evolution, and recently, there has been an effort to expand the realms in which entanglement can be shown to exist rigorously, particularly in ``natural'' systems -- i.e., not ones manufactured in laboratory conditions. Signatures of entanglement, a characteristically quantum feature, have been demonstrated in thermal states of bulk systems at low temperatures and between parties at macroscopic length scales \cite{Ved-2008}. Additionally, several recent studies have focused on the dynamics of entanglement in damped, driven, or generally non-equilibrium quantum systems \cite{Ved-2007, Qui.Rod.etal-2007, Cai.Pop.etal-2008, Tho.Eck.etal-2009}. The dynamics of entanglement in open systems can be extremely 
nontrivial -- especially in many-body systems -- and the precise influence of non-Hamiltonian dynamics on entanglement is poorly understood. In a result particularly relevant to this work, it is shown in Ref. \cite{Cai.Pop.etal-2008} that entanglement can be continuously generated and destroyed by non-equilibrium effects in an environment where no static entanglement exists. The possibility of entanglement in noisy non-equilibrium systems at high temperatures intimates the question: can we observe entanglement in the complex non-equilibrium chemical and biological processes necessary for life? Here we present strong evidence for answering this question in the affirmative by determining the timescales and temperatures for which entanglement is observable in a protein structure that is central to photosynthesis by green anoxygenic bacteria. 

\section{Light harvesting complexes and entanglement}
\label{sec:formal}
Recent ultrafast spectroscopic studies have revealed the presence of quantum coherence at picosecond 
 timescales in biological structures, specifically, in light harvesting complexes \cite{Eng.Cal.etal-2007, Lee.Che.etal-2007, Col.Won.etal-2010, Pan.Hay.etal-2010}. These studies demonstrate that in moderately strongly coupled, non-equilibrium systems, quantum features can be observed even in the presence of a poorly controlled, decohering environment. During the initial stage of photosynthesis, light is captured by pigment-protein antennas, known as light harvesting complexes, and the excitation energy is then transferred through these antennas to reaction centers where photosynthetic chemical reactions are initiated. Different LHCs vary in their detailed structure but all consist of densely packed units of pigment molecules, and all are spectacularly efficient at transporting excitation energy in disordered environments \cite{Bla-2002}. Average inter-chromophore separations on the scale of $\sim 15 \mathrm{\AA}$ are fairly common in LHCs. At these distances, the dipole coupling of these molecules is considerable and leads to coherent interactions at observable timescales \cite{Pul.Cha.etal-1996, Mon.Abr.etal-1997, Ame.Val.etal-2000}. It is this ``site'' coherence (coherence between spatially separated pigment molecules) that prompts us to examine entanglement in these systems and to consider the timescales and temperatures at which entanglement can exist. 

We begin by noting that in natural conditions many LHCs contain at most one excitation at any given 
time \cite{Bla-2002}. This is especially true of light harvesting complexes in photosynthetic bacteria, which are among the ones most heavily studied, because these bacteria receive very little sunlight in their natural habitat. 
Given this, we can treat each chromophore as a two-level system and the natural Hilbert space of the LHC quantum states will further be restricted to the zero and one excitation subspaces of the full tensor product space derived from $N$ chromophores.   Furthermore, there is no coherence between states in these different subspaces, because under natural conditions all processes connecting the two subspaces are incoherent. A state in the single-excitation manifold is written in the \textit{site basis} as:
\beq
\rho(t) = \sum_{i=1}^N \rho_{ii}(t) \ket{i}\bra{i} + \sum_{i=1}^N \sum_{j>i}^N \rho_{ij}(t) \ket{i}\bra{j} + \rho_{ij}^* (t)\ket{j}\bra{i},
\eeq
where $\ket{i}$ represents the state where only the $i$th chromophore (site) is excited and all other chromophores are in their electronic ground states. The density matrix $\rho$ is unnormalized -- that is, $\sum_i \rho_{ii}(t) \leq 1$ -- because the single excitation state has a finite lifetime due to trapping by the reaction center complex and radiative decay of the excitation. 

Given an unnormalized density matrix $\rho$ representing a single excitation state, we wish to calculate the amount of entanglement in the state. Entanglement in the site basis refers to non-local correlations between the electronic states of spatially separated chromophores. This is analogous to the entanglement of electromagnetic modes carrying a single photon in the dual-rail representation \cite{Enk-2005}. Mixed-state, multipartite entanglement is notoriously difficult to quantify \cite{Hor.Hor.etal-2009}, but the single-excitation restriction allows us to formulate two useful and simple measures. First, for bipartite entanglement between two chromophores, $i$ and $j$, we use a standard measure of entanglement, concurrence \cite{Hil.Woo-1997}, which, if $\rho$ is the single-excitation density matrix for the whole complex, takes the form $C_{ij} = 2|\rho_{ij}|$. Second, to quantify global entanglement in the LHC we derive (see Methods Section) a readily computable expression for mixed state entanglement in the physically relevant zero and single excitation subspaces, namely
\beq
E[\rho] = - \sum_{i=1}^N \rho_{ii}\ln \rho_{ii} - S(\rho),
\label{eq:measure}
\eeq
where $S(\rho) = -\tr \rho \ln \rho$ is the von Neumann entropy of the state $\rho$. In the Supplementary Information we show that this measure, which is based on the relative entropy of entanglement \cite{Ved.Ple.etal-1997}, is a true entanglement monotone over the physically relevant zero and single excitation states and hence provides a natural quantitative measure of entanglement in LHCs. 

\section{Entanglement dynamics in the FMO complex}
\label{sec:fmo}
A commonly studied LHC is the Fenna-Matthews-Olson (FMO) protein from green sulfur bacteria, such as \textit{Chlorobium tepidum} \cite{Cam.Bla.etal-2003}. The FMO complex is a trimer formed by three identical monomers that each bind seven bacteriochlorophyll-\textit{a} (BChl\textit{a}) molecules. We will restrict our study to a single monomer, which is shown in Fig. 1, since the monomers function independently. The site energies of the BChl\textit{a} molecules and coupling between molecules are well characterized; we use the \textit{Chlorobium tepidum} site energies and coupling strengths from Ref. \cite{Ado.Ren-2006} to form a Hamiltonian that describes the closed-system dynamics of an FMO monomer excitation (see the Supplementary Information for details). The structure of the Hamiltonian indicates that some pairs of chromophores are moderately strongly coupled (due to their close proximity and favorable dipole orientations) and hence effectively form \textit{dimers}. The wavefunctions of the system's energy eigenstates, usually called
 \textit{Frenkel excitons}, are primarily delocalized across these dimers. The dominant dimers are formed by chromophore pairs: 1-2, 5-6, and 3-4.

\begin{figure}[h]
\includegraphics[scale=0.4]{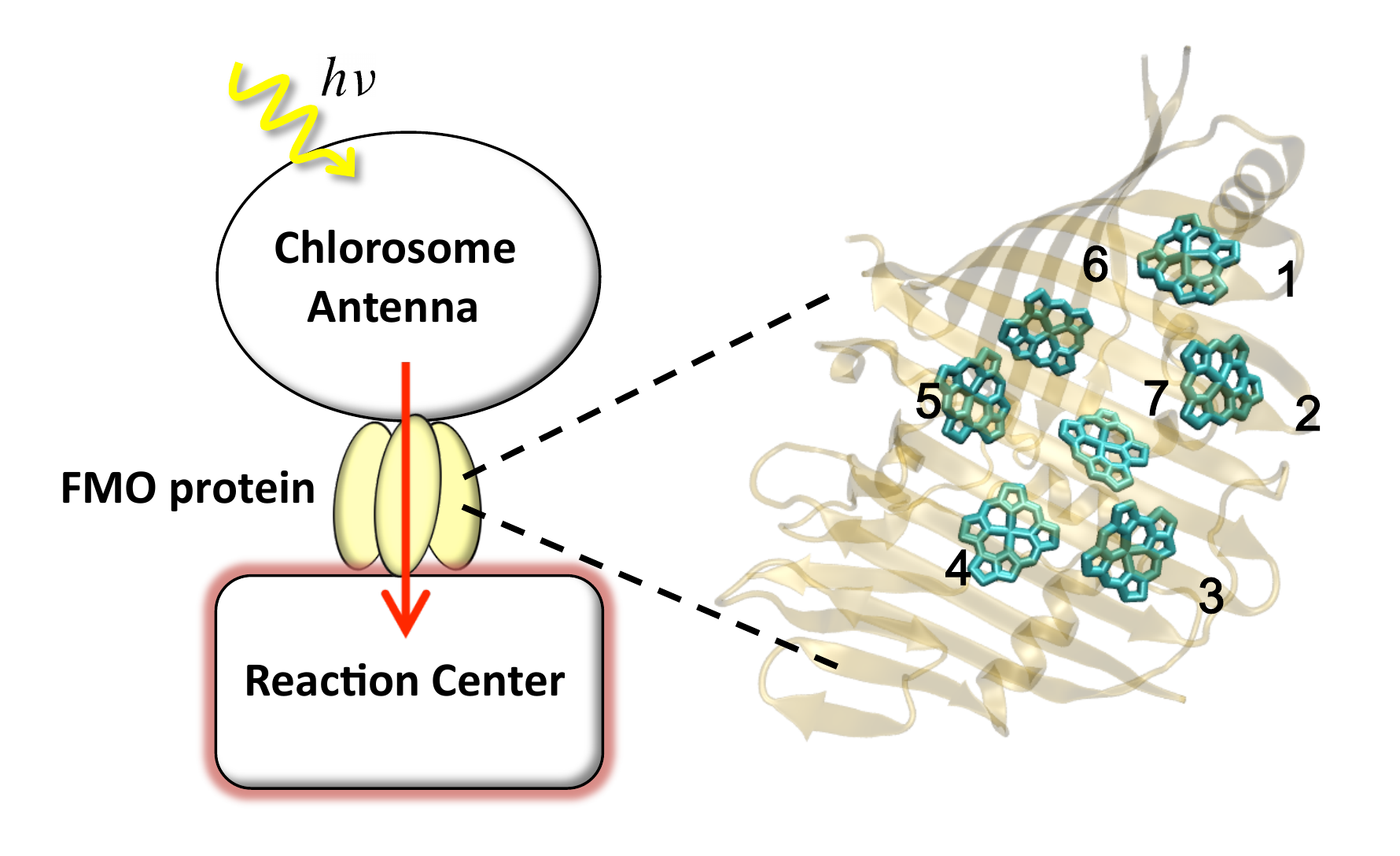}
\caption{\label{fig:fmo} The light harvesting apparatus of green sulfur bacteria and the Fenna-Matthews-Olson (FMO) protein. The schematic on the left illustrates the absorption of light by the chlorosome antenna and transport of the resulting excitation to the reaction center through the FMO protein. On the right is an image of a monomer of the FMO protein, showing also its orientation relative to the antenna and the reaction center \cite{Ado.Ren-2006,Wen.Zha.etal-2009}. The multi-ring units are bacteriochlorophyll-\textit{a} (BChl\textit{a}) molecules and the surrounding beta sheets and $\alpha$-helices form the protein environment in which the BChl\textit{a} molecules are embedded. The numbers label individual BChl\textit{a} molecules, also referred to as ``sites" in the main text. }
\end{figure}

In addition to the reversible (Hamiltonian) dynamics, there are interactions between each chromophore and the protein environment that it is embedded in. These interactions couple the protein dynamics to the FMO energy levels, resulting in static and dynamic disorder, and thereby lead to dephasing of the FMO electronic excitation state. Coherence properties of the FMO protein are then dictated by the interplay between coherent dynamics of the complex and decoherence effects due to environmental interactions. Distinctly quantum properties such as entanglement rely critically on coherence and therefore it is essential that a model that accurately accounts for environmental effects on coherence be used to make predictions about entanglement in FMO. In this work we use a recently 
developed non-perturbative, non-Markovian quantum master equation \cite{Ish.Fle-2009} to simulate excitation dynamics in the FMO complex. This dynamical model is particularly suited to describing the complex exciton-phonon interactions in light harvesting complexes and incorporates the dynamics of phonon reorganization in a realistic manner. See the Methods section for a brief review and discussion of this model, and the physical parameters we use in the simulations presented below.

\begin{figure}[t]
\includegraphics[scale=0.35]{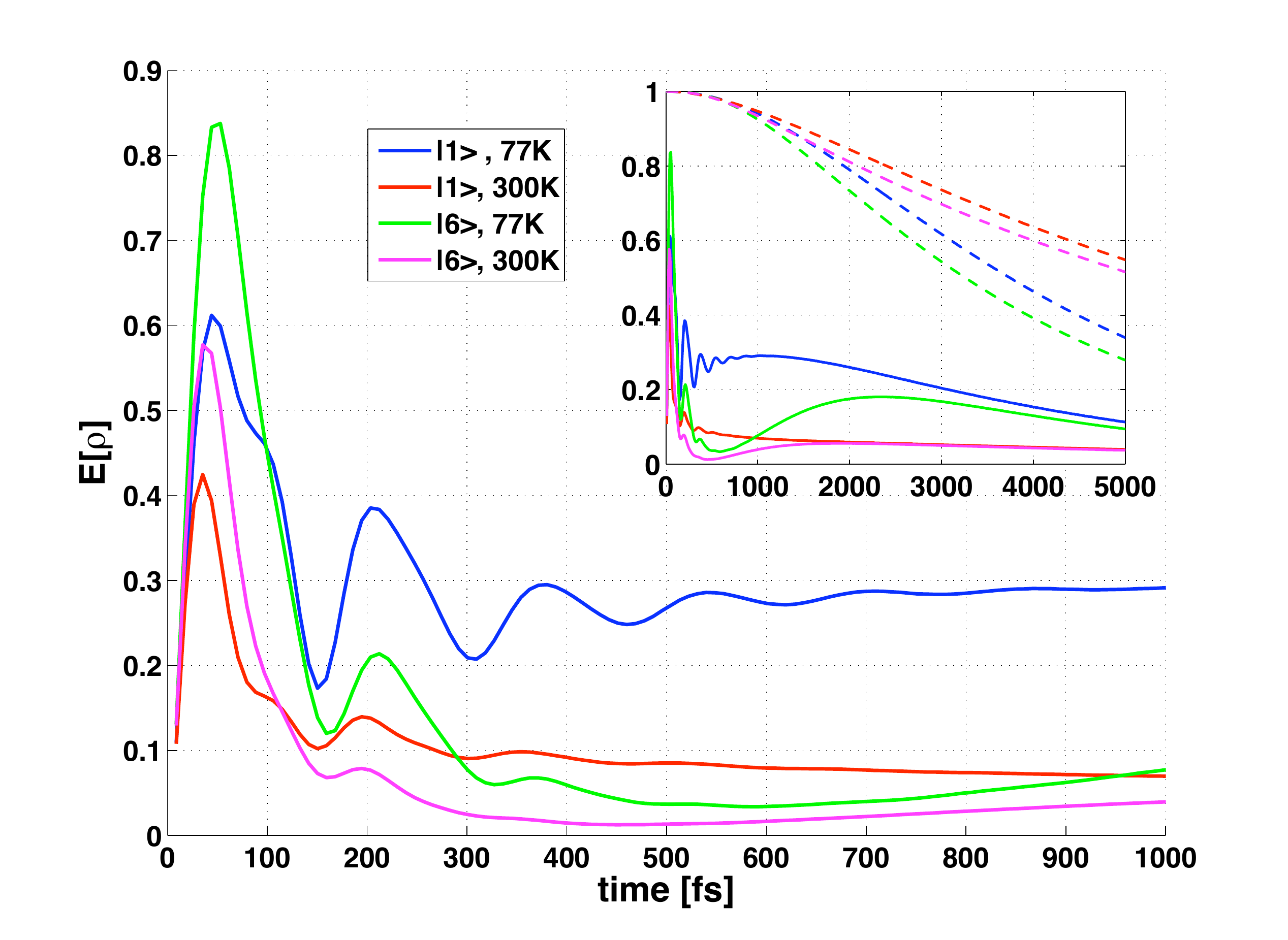}
\caption{\label{fig:E} Global entanglement in FMO. Time evolution of the global entanglement measure given in \erf{eq:measure} for the two initial states $\ket{1}$ and $\ket{6}$, at low ($T=77\,\mathrm{K}$) and high ($T=300\,\mathrm{K}$) temperatures. The inset shows the long-time evolution of the same quantities, together with the trace of the single excitation density matrix as dashed curves (identical color coding, and same units on axes as main figure).}
\end{figure}

Analysis of the FMO energy levels and its physical structure indicates that sites 1 and 6 interface with the chlorosome transmitting energy to the complex \cite{Ado.Ren-2006} (see Fig. 1), and recent experiments have confirmed that these two sites are the first to become excited in FMO \cite{Wen.Zha.etal-2009}. In Fig. 2 we show the time evolution of the global measure of entanglement given by \erf{eq:measure} when the initial state is an excitation on site $1$ or $6$, for two temperatures: 77\,K and 300\,K. In addition to simulations at the physiologically relevant temperature of 300\,K, we also perform simulations at 77\,K because the ultrafast spectroscopy experiments that probe LHCs are commonly performed at this temperature \cite{Eng.Cal.etal-2007, Lee.Che.etal-2007} (although very recently room temperature experiments have also been performed \cite{Col.Won.etal-2010, Pan.Hay.etal-2010}). 
The inset to this graph also includes the trace of the single excitation density matrix to show the total population in the single excitation subspace. A general feature of the global entanglement measure in all the scenarios depicted in Fig. 2 is that its value rises rapidly for short times and then after $\sim 30-50\,\mathrm{fs}$ decays with varying amounts of oscillation. An explanation for this behavior is that entanglement increases rapidly for short times due to quick delocalization of the excitation caused by large $1$-$2$ and $5$-$6$ site coupling terms in the FMO complex Hamiltonian. Then, as the excitation begins exploring other sites, the global entanglement decreases due to incoherent transport and rapid dephasing. For both initial conditions and temperatures, there is finite entanglement in the system up to $5$ps. At 300\,K the short time behavior is qualitatively similar to the low temperature case. At long times, however, the initial state dependence of the global entanglement is suppressed in contrast to the low temperature case. 

To further elucidate the dynamics and structure of entanglement in the FMO monomer under realistic conditions, we examine pairwise entanglement in the system, as measured by the bipartite concurrence between two sites: $C_{ij} = 2|\rho_{ij}|$ for any two sites $i, j$. Figures 3 and 4 show the time evolution of pairwise concurrence when the initial state is $\ket{1}$ and $\ket{6}$, respectively, at both 77\,K and 300\,K. For clarity, we have only shown the most significant and illustrative concurrence curves, but it should be noted that there is finite bipartite entanglement between almost all chromophores, especially for short time scales. The Supplementary Information section contains plots that provide a complete depiction of bipartite entanglement in this complex.

When the initial excitation is on site 1, as shown in Fig. 3, the bipartite entanglement in the complex is primarily between sites $1, 2, 3$ and $4$. Most of this bipartite entanglement is between pairs of moderately strongly coupled chromophores that form dimers and hence have large amounts of coherence. However, a surprising aspect of the short time behavior is that there is considerable concurrence between non-nearest-neighbor sites, as indicated by the $1$-$5$ and $1$-$3$ curves. Perhaps most strikingly, there is a large amount of entanglement within 1\,ps between chromophores 1 and 3, which are almost the furthest apart in the FMO complex (separated by $\sim 28 \mathrm{\AA}$), and are very weakly coupled. This long-range entanglement is mediated by chromophores connecting these sites, and is a sign of multipartite entanglement in the system. At 77\,K bipartite entanglement persists for greater than 5\,ps, and seems limited only by the energy trapping rate -- essentially there is non-negligible entanglement as long as the excitation has not been trapped by the reaction center. Bipartite entanglement persists for long periods of time at 300\,K also, but it diminishes at a faster rate compared to low temperatures. However, note that at short times ($<600\,\mathrm{fs}$) temperature has little effect on the amount of bipartite entanglement in the system -- qualitatively, the concurrence is scaled by $\sim 3/4$ when the temperature is increased from 77\,K to 300\,K. 

\begin{figure}[t]
\includegraphics[scale=1]{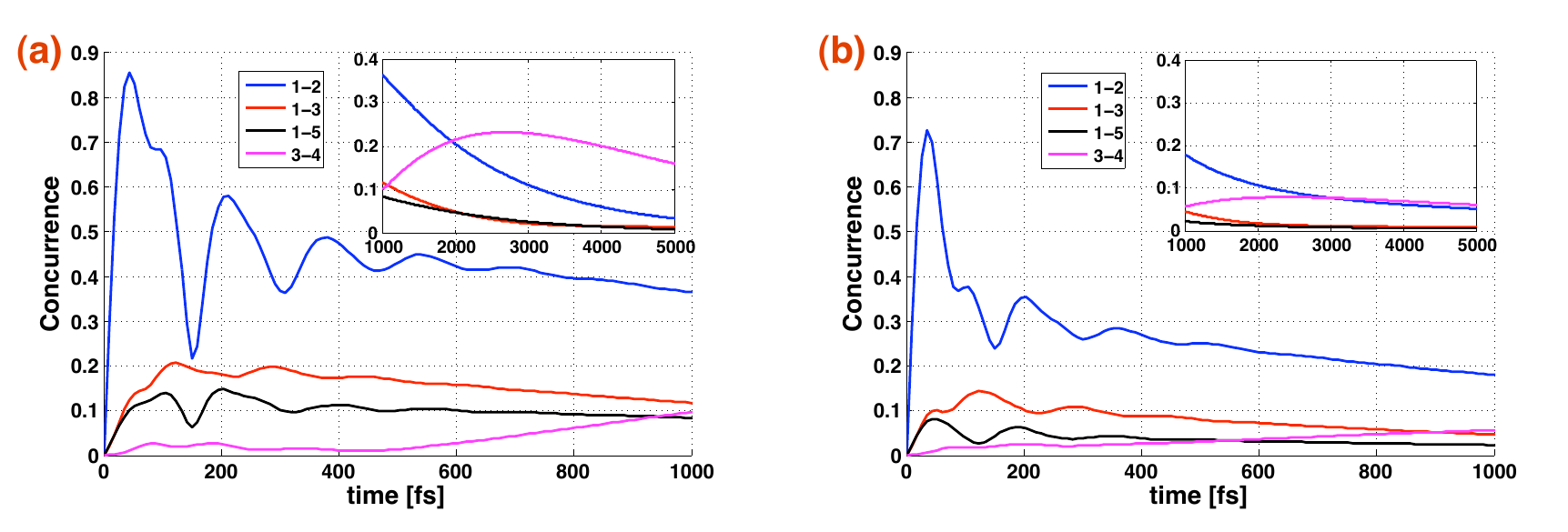}
\caption{\label{fig:conc_1} Bipartite entanglement in FMO when initial state is an excitation localized on site 1. The figure shows time evolution of the concurrence measure of bipartite site entanglement, $C_{ij}=2|\rho_{ij}|$, at (a) 77\,K, and (b) 300\,K. Only curves for the most entangled chromophores are shown. Insets show the long time behavior (identical color coding, and same units on axes as main figures). }
\end{figure}

When the initial excitation is on site 6, as shown in Fig. 4, bipartite entanglement exists between several chromophores at short times; there is non-negligible entanglement between any two of the sites: $4, 5, 6$ and $7$. Note that not all of these are dimerized chromophores with large couplings. Furthermore, the presence of finite entanglement across any bipartite partition of the sites $4,5,6$ and $7$ indicates that the FMO complex contains genuine multipartite distributed entanglement within $\sim 600\,\mathrm{fs}$. 

\begin{figure}[t]
\includegraphics[scale=1]{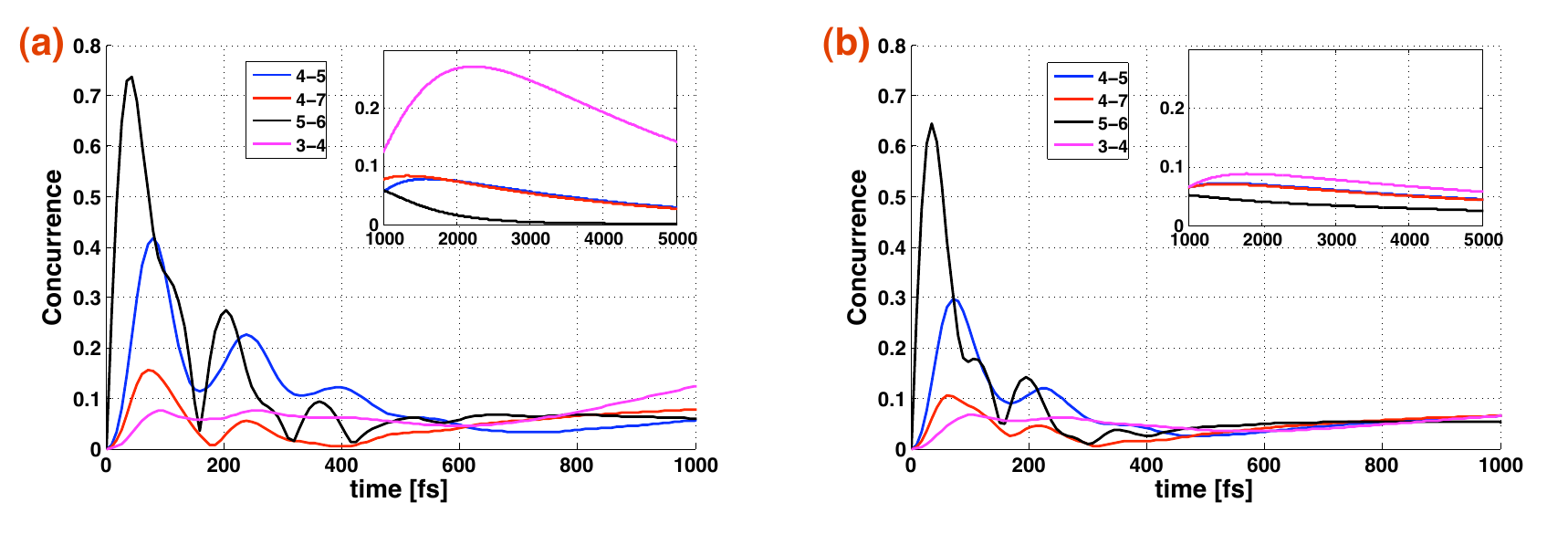}
\caption{\label{fig:conc_6} Bipartite entanglement in FMO when initial state is an excitation localized on site 6. The figure shows time evolution of the concurrence measure of bipartite site entanglement, $C_{ij}=2|\rho_{ij}|$, at (a) 77\,K, and (b) 300\,K. Only curves for the most entangled chromophores are shown. Insets show the long time behavior (identical color coding, and same units on axes as main figures).}
\end{figure}

To summarize, realistic simulations of FMO dynamics indicate that considerable multipartite entanglement is present in the FMO complex at timescales of $\sim 5\,\mathrm{ps}$ at 77\,K and $\sim 2\,\mathrm{ps}$ at 300\,K. For both realistic initial states there is multipartite entanglement between all sites involved in excitation transport.

\section{\label{sec:discussion} Significance and implications of LHC entanglement}

The previous section presented numerical evidence for the existence of entanglement in the FMO complex for picosecond timescales -- essentially until the excitation is trapped by the reaction center. This is remarkable in a biological or disordered system at physiological 
temperatures. It illustrates that non-equilibrium multipartite entanglement can exist for relatively long times, even in highly decoherent environments. While the length scales over which entanglement was shown to persist were restricted to $\lesssim 30 \mathrm{\AA}$ because of the relatively small size of the FMO complex, we expect that such long-lived, non-equilibrium entanglement will also be present in larger light harvesting antenna complexes, such as LH1 and LH2 in purple bacteria. This is because they contain the key necessary ingredient; moderately strongly coupled chromophores that can lead to significant coherent delocalization of electronic excitations \cite{Mei.Zha.etal-1997, Mon.Abr.etal-1997}. In larger light harvesting antennae it may also be possible to take advantage of the ability to create and support multiple excitations in order to access a richer variety of entangled states. 

Our numerical results are based on a non-perturbative, non-Markovian dynamical model particularly formulated to model the actual conditions in light harvesting complexes. The precise dynamics of entanglement at short-times is dependent on the non-Markovian dynamics of the system, and a correct picture of which chromophores are entangled at long times relies on an accurate approach to thermal equilibrium. The dynamical model we use realistically models both these limits. We note however, that the long-time survival of entanglement in the FMO complex is also predicted by less sophisticated dynamical models (see Supplementary Information) and therefore is a robust feature of these systems.

We emphasize that our prediction of entanglement in the FMO complex is experimentally verifiable because the timescales on which the entanglement exists can readily be probed using non-linear femtosecond spectroscopy techniques. The ability to perform quantum state tomography on light harvesting complexes is not currently possible, but techniques are under development that will allow one to extract individual density matrix elements. The primary obstacle to this task is the determination of excitonic transition dipoles in the complex because the spectroscopy signals representing elements of the density matrix are scaled by the magnitude of these dipoles. It has recently been demonstrated that polarization-dependent two-dimensional spectroscopy is capable of determining these dipole moments \cite{Rea.Sch.etal-2008}, and hence the dynamical estimation of density matrix elements is within reach. An entanglement witness, $\mathcal{W}$, identified in the Methods section simplifies the experimental verification of entanglement since it implies that monitoring a small subset of site basis coherences (for example, the ones expected to be the largest in magnitude) is sufficient for detecting entanglement.

It is interesting to consider the biological significance of this entanglement. The FMO complex is an unusual light harvesting component because its primary role is as a wire, to transport excitations between the main light harvesting complex of green sulfur bacteria and their reaction centers. Most other light harvesting structures have dual roles of light capture and excitation transport and are as a consequence larger and more complex. This study demonstrating the essential features of entanglement in LHCs focused on the FMO complex because it is very well characterized. While FMO may be too small to take advantage of the non-classical correlations implied by entanglement, we expect that such entanglement has functional implications in more complex light harvesting structures, particularly in photosynthetic units with multiple reaction centers. It is possible that strong correlations between distant sites may enhance properties of excitation transport (e.g. robustness, efficiency, regularity) through such light harvesting networks. Further study is required to determine the precise role the non-classical correlations that are realized by this entanglement play in light harvesting dynamics of larger systems. We note that a functional role for entanglement in LHCs is consistent with studies of the effects of inter-site electronic coherence on kinetic properties of LHCs that have found such coherence benefits efficiency \cite{Muk.Abe.etal-1999, Sch.Fle-2000, Sum-2001, Jan.New.etal-2004, Moh.Reb.etal-2008, Ple.Hue-2008} and robustness \cite{Che.Sil-2006} of excitation transport.

Our framework for quantification of entanglement in LHCs is an essential step in the precise characterization of quantum resources in organic structures, and is of particular relevance to the endeavour of constructing quantum devices from such structures. While coherence is a feature of quantum mechanical evolution whose quantification is not directly operationally material, the non-classical correlations embodied by entanglement constitute an operational resource. Entanglement resulting from delocalization of a single particle among experimentally accessible locations (e.g an excitation among the chromophores of an LHC) represents non-local correlations with compelling technological applications \cite{Enk-2005, Lom.Sci.etal-2002}. In particular, the presence of entanglement in LHCs sets the stage for investigating the applicability of entanglement-enhanced measurement \cite{Gio.Llo.etal-2004} in biological systems. Quantum metrology enabled by the use of entangled states has been demonstrated using photons and ions, and these engineered systems have been used to measure frequencies and phases to unprecedented precision (e.g. \cite{Roo.Chw.etal-2006, Nag.Oka.etal-2007}). The application of ideas from entanglement-enhanced metrology to the improved measurement of biological properties in these naturally robust quantum coherent systems is likely to be fruitful. Within this context, note that the observed enhanced radiative decay of delocalized electronic excitation states of LH1 and LH2 \cite{Mon.Abr.etal-1997}, 
which by the arguments of this paper are also entangled states in the single excitation subspace, is consistent with the fact that entangled states can be more sensitive probes of the environment than separable states.

In addition to quantum-enhanced metrology, densely packed molecular aggregates such as LHCs have potential for constructing naturally robust quantum devices. For example, ultra-fast quantum state transfer facilitated by excitation migration along engineered or self-assembled chromophoric arrays could be a possible realization of the ``quantum wires" that are much sought after in quantum technology. Entanglement between sub-units in such an array is essential for high-fidelity quantum transport, and as our study shows, such entanglement is possible in molecular aggregate structures even at room temperature. This conclusion is reinforced by recent studies at room temperature that have presented evidence for delocalization of excitations along molecular wires formed by self-assembling \textit{J}-aggregates \cite{Lag.Sou.etal-2004}. Integration of such molecular aggregates with solid-state devices (e.g. cavities \cite{Tis.Bra.etal-2007}) opens up the possibility of engineering controllable quantum coherent soft matter structures that can be used to distribute quantum states or entanglement. It should be noted that the control overhead for quantum state transfer in such wires is not overly prohibitive -- it has been shown that the only requirement for high fidelity state transfer in \textit{randomly} coupled linear chains is complete control (e.g optical addressability) over the end points of the linear chain \cite{Bur.Bos-2005}. 

We conclude by returning to the foundational significance of entanglement; it represents a uniquely quantum form of strong correlation between physical systems. The identification of entanglement -- which was referred to by Schr\"odinger as ``\textit{the} characteristic trait of quantum mechanics, the one that enforces its entire departure from classical lines of thought'' \cite{Sch-1935a} -- between spatially distinct components of a biological system under natural functioning conditions further expands the field of physical systems for which nontrivial and uniquely quantum signatures become manifest.

\section{Methods}
\label{sec:methods}

\subsection{Measures of entanglement in LHCs}
The single excitation assumption valid for many light harvesting complexes allows us to formulate simple measures of entanglement in these systems. First, note that if $\rho$ represents a single excitation state in the site basis, we have the following: 
\begin{quote}
\textbf{\textit{Proposition 1:}} $\rho$ entangled $\iff \rho_{ij} \neq 0$ for some $i\neq j$. 
\end{quote}
Here, the right hand side simply means that $\rho$ has some coherence. Hence, coherence (in the site basis) in the single excitation subspace is necessary and sufficient for entanglement. \\
\textbf{\textit{Proof:}} The forward implication of proposition 1 is clearly seen from its contrapositive form: $\rho_{ij}=0 ~~\forall i\neq j \Longrightarrow \rho$ separable. If $\rho_{ij}=0 ~~\forall i\neq j$, then the state has the form $\rho = \sum_i \rho_{ii} \ket{i}\bra{i}$, which is clearly a separable state since each $\ket{i}$ is separable. To prove the backward implication, again consider its contrapositive form: $\rho$ separable $\Longrightarrow \rho_{ij}=0 ~~\forall i\neq j$. If $\rho$ is separable it contains no entanglement, and specifically no bipartite entanglement between any two chromophores. The well known mixed state bipartite entanglement measure for two-level systems, \textit{concurrence} \cite{Hil.Woo-1997}, can be computed between chromophores $1$ and $2$. In the single excitation subspace, it evaluates to $C_{12} = 2|\rho_{12}|$. Hence, zero bipartite entanglement between these chromophores implies $\rho_{12}=0$. Similarly, by setting the bipartite entanglement between all pairs of chromophores to zero, we find that $\rho_{ij}=0 ~~\forall i\neq j.$

This property of states in the single excitation subspace implies that an effective entanglement witness \cite{Hor.Hor.etal-1996, Per-1996} for these states is simply the sum of all coherences: 
\beq
\mathcal{W} = \sum_{i<j} |\rho_{ij}|
\label{eq:witness}
\eeq
That is, $\mathcal{W}>0 \Longleftrightarrow \rho$ entangled. Note that unlike most entanglement witnesses, which only present sufficient conditions for entanglement, in this case $\mathcal{W}>0$ is both necessary and sufficient for entanglement. Practically this property can therefore be used in experiments to detect entanglement; any non-zero off-diagonal component of the LHC single excitation density matrix in the site basis is a signature of entanglement. Simplifications of this witness, which become sufficient conditions for entanglement, can also be formulated by restricting the sum to be over a subset of the off-diagonal elements -- this could be useful in situations where accessing all off-diagonal elements is experimentally prohibitive. 

Lastly, we can also construct a measure of global entanglement in this system based on the relative entropy of entanglement, which is defined as \cite{Ved.Ple.etal-1997}: 
\beq
E[\rho] = \min_{\sigma \in \mathcal{D}} S(\rho || \sigma)
\eeq
where $\mathcal{D}$ is the set of separable states and $S$ is the relative entropy function: $S(\rho || \sigma) = \tr(\rho \ln \rho - \rho \ln \sigma)$. Since the physically relevant Hilbert space for LHCs is restricted to the zero and single excitation subspaces, we define the natural photosynthetic entanglement to be the \emph{restricted relative entropy of entanglement} in which the minimization is performed over $\sigma \in \mathcal{D}^{*}$, the set of separable states in the zero and single excitation manifold. We know from \textit{Proposition 1} above that the set of separable states in this restricted Hilbert space has diagonal form. Consequently the minimization problem becomes:
\bqa
E[\rho] = \min_{p_i} \tr( \rho \ln \rho &-& \rho \ln \sigma), \nn \\
\textrm{subject to the constraint} ~&&~ \sum_{i=1}^N p_i = \tr\rho,
\eqa
where $\sigma = \textrm{diag}(p_1, p_2, ..., p_N)$ and the optimization constraint derives from the fact that we are only considering separable states with the same normalization as $\rho$. Since the cost function $E[\rho]$ is convex in $p_i$ and the constraint is linear, this optimization is easily solved, e.g., by using a Lagrange multiplier to combine the cost function and constraint into a Lagrangian and then finding its stationary point. This results in an explicit expression for the closest separable state, namely $\sigma^* = \textrm{diag}(p_1^*, p_2^*, ..., p_N^*)$ with $p_i^* = \rho_{ii}$ as well as the following measure of global entanglement in the state $\rho$:
\beq
E[\rho] = - \sum_{i=1}^N \rho_{ii}\ln \rho_{ii} - S(\rho),
\eeq
 with $S(\rho) = -\tr \rho \ln \rho$ the von Neumann entropy of $\rho$. We note that, consistent with our discussion above, $E[\rho]$ is also a measure of coherence in the system, since the first term on the right hand side is the entropy of a state for which all coherences $\rho_{ij}$ are artificially set to zero while the von Neumann entropy implicitly contains all coherences. We show in the Supplementary Information that this photosynthetic entanglement measure satisfies the properties of an entanglement monotone \cite{Hor.Hor.etal-2009} within the restricted zero and single excitation Hilbert space.

\subsection{Reduced hierarchy equations approach to modeling excitation dynamics in LHCs}

Electronic excitation dynamics is governed by a Hamiltonian of the form: $H = H_{\textrm{el}} + H_{\textrm{el-env}} + H_{\textrm{env}}$, where 
\bqa
H_{\textrm{el}} &=& \sum_{i=1}^N E_i \ket{i}\bra{i} + \sum_{i=1}^N \sum_{j>i}^N J_{ij}(\ket{i}\bra{j} + \ket{j}\bra{i})
\eqa
describes the closed system dynamics of the $N$ LHC chromophores, including on-site energies, $E_j$, and coupling terms, $J_{ij}$, that describe the coupling between the chromophores. The remaining terms in $H$ describe the coupling between the electronic degrees of freedom of the LHC chromophore molecules and their environment, which typically consists of surrounding proteins, the electromagnetic field, and reaction center/s that the LHC is affixed to. The dominant environmental perturbations are the 
phonons of the protein scaffolding around the LHC, and these are modeled as a diagonal coupling to a phonon bath:
\beq
H_{\textrm{el-env}} = \sum_{i=1}^{N} \ket{i}\bra{i} \sum_{\xi} g^{i}_{\xi} Q^{i}_{\xi}
\label{eq:bilinear}
\eeq
where $Q^{i}_{\xi}$ is a phonon mode indexed by $\xi$, which is coupled to chromophore $i$ with coupling strength $g^{i}_{\xi}$. Obtaining a faithful reduced description of the effective dynamics of just the electronic excitation state $\rho$ requires a physically accurate averaging over the external degrees of freedom and quantum master equation approaches are commonly utilized for this averaging. In particular, the standard Redfield equation \cite{Red-1957, May.Kuh-2004} is often used to explore the dynamics of photosynthetic excitation energy transfer (EET). This equation is valid when the exciton-phonon coupling -- which can be specified by the magnitude of the reorganization energy -- is much smaller than the interchromophoric coupling because the equation is derived on the basis of a second-order perturbative truncation with respect to the exciton-phonon coupling. However in many LHCs, the reorganization energies are not small in comparison to the interchromophoric coupling: \textit{e.g.} in FMO the electronic coupling strengths span a wide range, $1\sim100\,\mathrm{cm}^{-1}$, and the reorganization energies span a similar range \cite{Bri.Ste.etal-2005, Cho.Vas.etal-2005, Ado.Ren-2006, Rea.Sch.etal-2008}. Hence, the standard Redfield equation approach might lead to erroneous insights and incorrect conclusions regarding quantum coherent effects in the FMO complex \cite{Ish.Fle-2009a}. 
Two of the present authors have recently presented a reliable theoretical framework to describe photosynthetic EET that takes into account the phonon relaxation dynamics associated with each chromophore in chromophore-protein complexes \cite{Ish.Fle-2009}.  This framework can describe quantum coherent wave-like motion and incoherent hopping motion in a unified manner, and reduces to the standard Redfield theory and F\"orster theory \cite{For-1948, May.Kuh-2004} in their respective limits of validity. The ability of the framework to interpolate between these two limits is significant because photosynthetic EET commonly occurs between these two perturbative regimes, as in the case of the FMO complex. The only assumptions used in this framework are that: (i) the exciton-phonon coupling is bilinear (e.g. \erf{eq:bilinear}),  (ii) the environmental fluctuations are described as Gaussian processes, (iii) the total system is in a factorized (product) initial state, and (iv) an overdamped Brownian oscillator model for the phonon environment which results in exponentially (time) correlated phonon fluctuations.
The hierarchical expansion technique \cite{Tan-2006} is employed in order to obtain a practical expression for numerical calculations \cite{Ish.Fle-2009}.

For the simulations in this work we take the reorganization energy of the molecular environment to be $35\,\textrm{cm}^{-1}$, a value that is consistent with experimentally extracted reorganization energies for this complex \cite{Bri.Ste.etal-2005, Cho.Vas.etal-2005, Ado.Ren-2006, Rea.Sch.etal-2008}. 
We choose a phonon relaxation time of 100\,fs, and a reaction center trapping rate associated with site 3 of $(4\,\textrm{ps})^{-1}$, both of which are consistent with the literature on LHCs \cite{Cho.Vas.etal-2005, Ado.Ren-2006, Ple.Hue-2008}.
We do not assume any form of spatial correlations in the phonon fluctuations. This is primarily because there is no experimental data on spatially correlated phonon fluctuations in FMO and the goal here is for the simulations to be as close as possible to what is currently known about the FMO environment. We note that at short times, spatially correlated fluctuations are expected to increase coherence in the site basis, and hence will naturally increase the entanglement present at initial times. Therefore the predictions in this work, which are based on simulations that assume no spatial correlations in the environment, could be viewed as a lower bound on the amount of entanglement in the FMO complex.

Finally, we have compared the entanglement predicted by the 
present framework to that predicted by standard Markovian Redfield models. This comparison is presented in the Supplementary Information.

\section{Supplementary Information}
\subsection{Photosynthetic entanglement measure}

The global entanglement measure for light harvesting complexes that is presented in the main text and derived in the Methods Section can be proven to be a true entanglement monotone within the physically relevant Hilbert space spanned by zero and single excitation states. Recall that the photosynthetic entanglement measure takes the form:
\beq
E[\rho] = -\sum_{i} \rho_{ii}\log \rho_{ii} - S(\rho).
\label{eq:meas}
\eeq
For convenience we define $\mathcal{P}$ as a projector onto the subspace of zero and one excitation states of the full Hilbert space, $\mathbb{C}^{2N}$.

In order to compare the amount of entanglement in different states it is desirable to have a measure that is a monotonic function of local transformations \cite{Vid-2000}.   An entanglement monotone should satisfy the following four properties \cite{Wei.Nem.etal-2003}:

\begin{enumerate}
\item $E[\rho] \geq 0$, with equality if $\rho$ is separable. \\
\item $E[\rho]$ is invariant under local unitary operations. That is, if $U_{L}= U_{1}\otimes U_{2}\otimes ... \otimes U_{N}$, $E[U_{L}\rho U_{L}\dg] = E[\rho]$. \\
\item Local operations or classical communication do not increase $E[\rho]$. \\
\item $E[\rho]$ is convex under loss of information -- i.e. $\sum_{i}p_{i} E[\rho_{i}] \geq E\left[\sum_{i}p_{i}\rho_{i} \right]$. \\
\end{enumerate}

We now prove that the measure of natural photosynthetic entanglement defined by Eq.~(\ref{eq:meas}) satisfies each of these conditions and thus is a true entanglement monotone.

\begin{enumerate}
\item $E[\rho] \geq 0$, with equality if $\rho$ is separable. \\
\emph{Proof:} The first quantity in \erf{eq:meas} is the entropy of the state $\rho$ after it has been projected onto the site basis. That is, if $\bar{\rho} = \sum_{i=1}^{N} \bra{i}\rho \ket{i} \ket{i}\bra{i}$, then $E[\rho] = S(\bar{\rho}) - S(\rho)$. One of the properties of von Neumann entropy is that $S(\bar{\rho}) \geq S(\rho)$ \cite{Bre.Pet-2002}. Hence $E[\rho] \geq 0$. It is also clear that $E$ is zero for all separable states invariant under $\mathcal{P}$, because these separable states are all diagonal.

\item $E[\rho]$ is invariant under local unitary operations. That is, if $U_{L}= U_{1}\otimes U_{2}\otimes ... \otimes U_{N}$, $E[U_{L}\rho U_{L}\dg] = E[\rho]$. \\
\emph{Proof:} The invariance of $E$ under local unitaries follows from the invariance of the relative entropy under unitary operations \cite{Ved-2002}. $E$ is simply a relative entropy and since $U_{L}$ does not take a state out of the zero and single excitation subspace, then:
\bqa
\min_{\sigma \in \mathcal{PD}} S(U_{L} \rho U_{L}\dg || \sigma) =  \min_{\sigma \in \mathcal{PD}} S(U_{L} \rho U_{L}\dg || U_{L} \sigma U_{L}\dg) = \min_{\sigma \in \mathcal{PD}} S(\rho || \sigma) \equiv E[\rho]
\eqa
where the subspace of zero and single excitation separable states is denoted $\mathcal{P}\mathcal{D}$. The first equality comes from the fact that if $\mathcal{P}U_{L}\mathcal{P} = U_{L}$, then the set of zero and single excitation separable states is mapped into itself by $U_{L}$. The second equality comes from the unitary invariance of relative entropy  \cite{Ved-2002}.

\item Local operations or classical communication do not increase $E[\rho]$. \\
\emph{Proof:} This follows in the same way as the previous proof. That is, let $\mathcal{O}_{L}$ denote the local operation and classical communication (a general local completely positive map). Then,
\beq
\min_{\sigma \in \mathcal{PD}} S(\mathcal{O}_{L}(\rho) || \sigma) =  \min_{\sigma \in \mathcal{PD}} S(\mathcal{O}_{L}(\rho) || \mathcal{O}_{L}(\sigma)) \leq \min_{\sigma \in \mathcal{PD}} S(\rho || \sigma) \equiv E[\rho]
\eeq
The inequality results from the non-increasing property of relative entropy under any completely positive map \cite{Ved-2002}. Note that the property $\mathcal{P}\mathcal{O}_{L}(\sigma)\mathcal{P} = \mathcal{O}_{L}(\sigma)$ is necessary for the first equality. 

\item $E[\rho]$ is convex under loss of information -- i.e. $\sum_{i}p_{i} E[\rho_{i}] \geq E\left[\sum_{i}p_{i}\rho_{i} \right]$. \\
\emph{Proof:} This property is proved for the relative entropy of entanglement in Ref. \cite{Ved.Ple-1998}. The restriction of the minimization to the zero and single excitation subspace does not effect this property (as long as all $\rho_{i}$ are in the restricted subspace). 
\end{enumerate}

\subsection{Complete characterization of bipartite entanglement}
In the main text we presented the dynamics of bipartite entanglement for selected chromophore pairs. These were selected to elucidate the longevity and multipartite nature of the entanglement in FMO. Here we show the dynamics of bipartite entanglement, as measured by the concurrence $C_{ij} = 2|\rho_{ij}|$, for all pairs of chromophores in the complex, except those showing negligible entanglement ($C_{ij}<0.05$ for all time).

\begin{figure}[h]
\centering
\includegraphics[width=18cm]{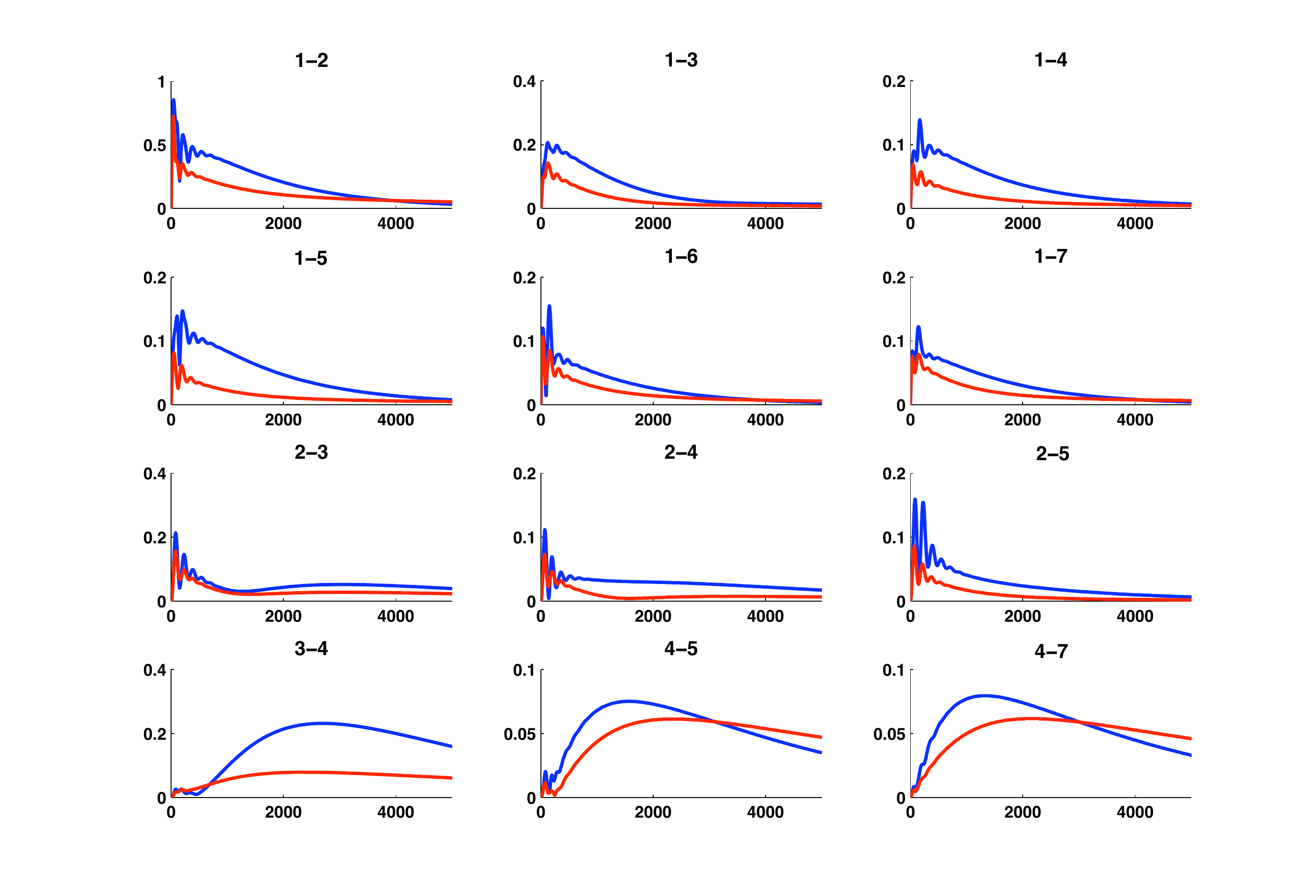}
\caption{Bipartite entanglement when the initial state is $\ket{1}$. All vertical axes show concurrence, $C_{ij}=2|\rho_{ij}|$, and all horizontal axes show time in units of femtoseconds. The blue curves are for T=77K, and the red curves are for T=300K.
	\label{fig:conc_full_1} }
\end{figure}

\begin{figure}[h]
\centering
\includegraphics[width=18cm]{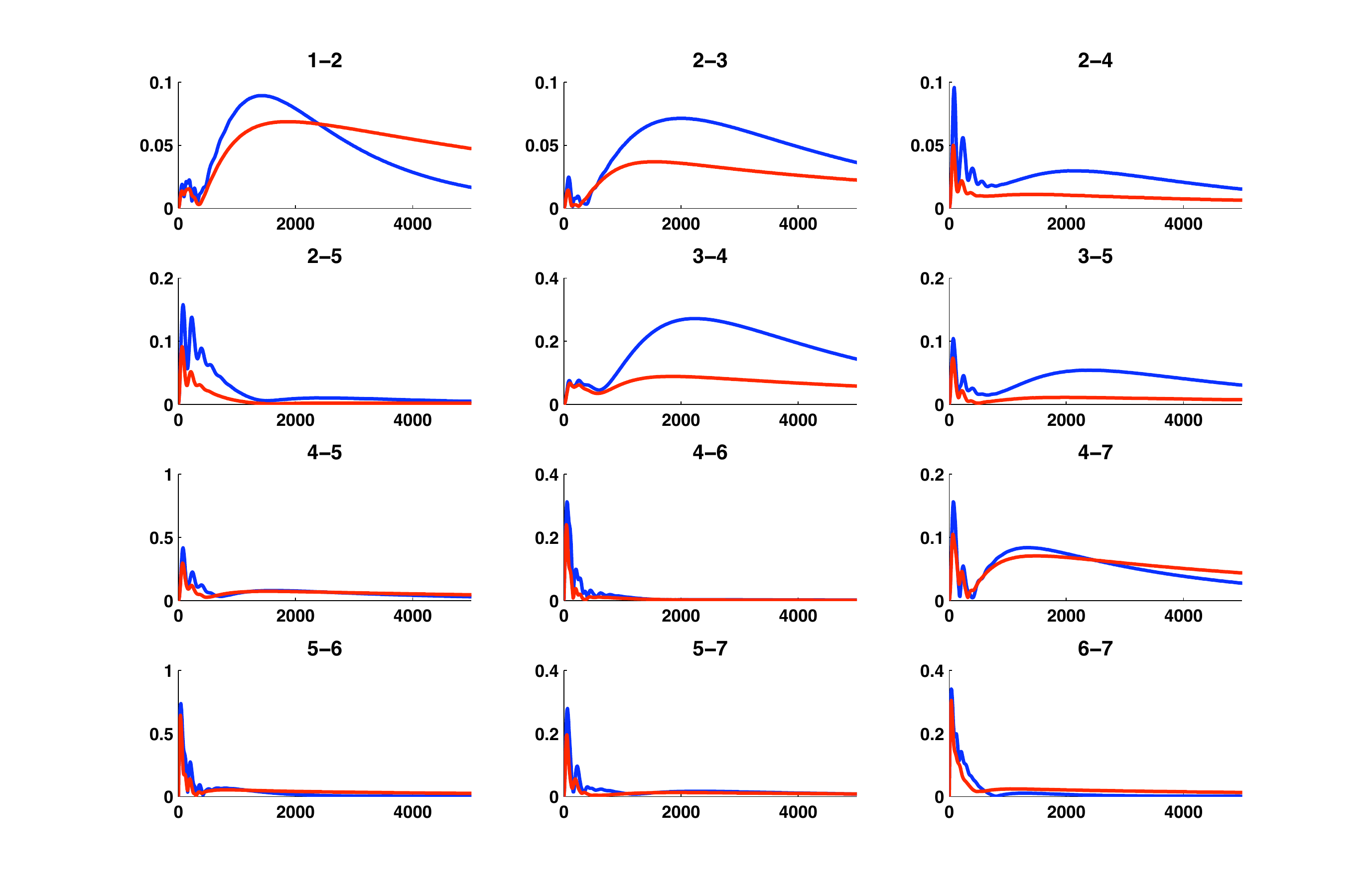}
\caption{ Bipartite entanglement when the initial state is $\ket{6}$. All vertical axes show concurrence, $C_{ij}=2|\rho_{ij}|$, and all horizontal axes show time in units of femtoseconds. The blue curves are for T=77K, and the red curves are for T=300K.
	\label{fig:conc_full_6} }
\end{figure}

Fig. \ref{fig:conc_full_1} shows concurrence versus time for two temperatures when the initial state is $\ket{1}$. Fig. \ref{fig:conc_full_6} shows concurrence versus time for two temperatures when the initial state is $\ket{1}$. 

\subsection{Comparison of dynamical models}

Here we compare the predictions for entanglement in the FMO complex made from an analysis of the simulations with realistic incorporation of protein reorganization dynamics that are presented in the main text of the paper, to predictions made from analyses of more standard simulations based on the Markovian Redfield equation. Two versions of Redfield theory are commonly used: the Redfield model with the secular approximation (which is equivalent to the Lindblad formulation of Markovian dynamics \cite{Lin-1975, Bre.Pet-2002}), and the full Markovian Redfield equations with no additional approximations. Redfield dynamics have recently been used to study the interplay between coherent dynamics and dephasing on excitation energy transfer in FMO \cite{Ple.Hue-2008, Reb.Moh.etal-2009}.

\begin{figure}[t]
\includegraphics[width=9cm]{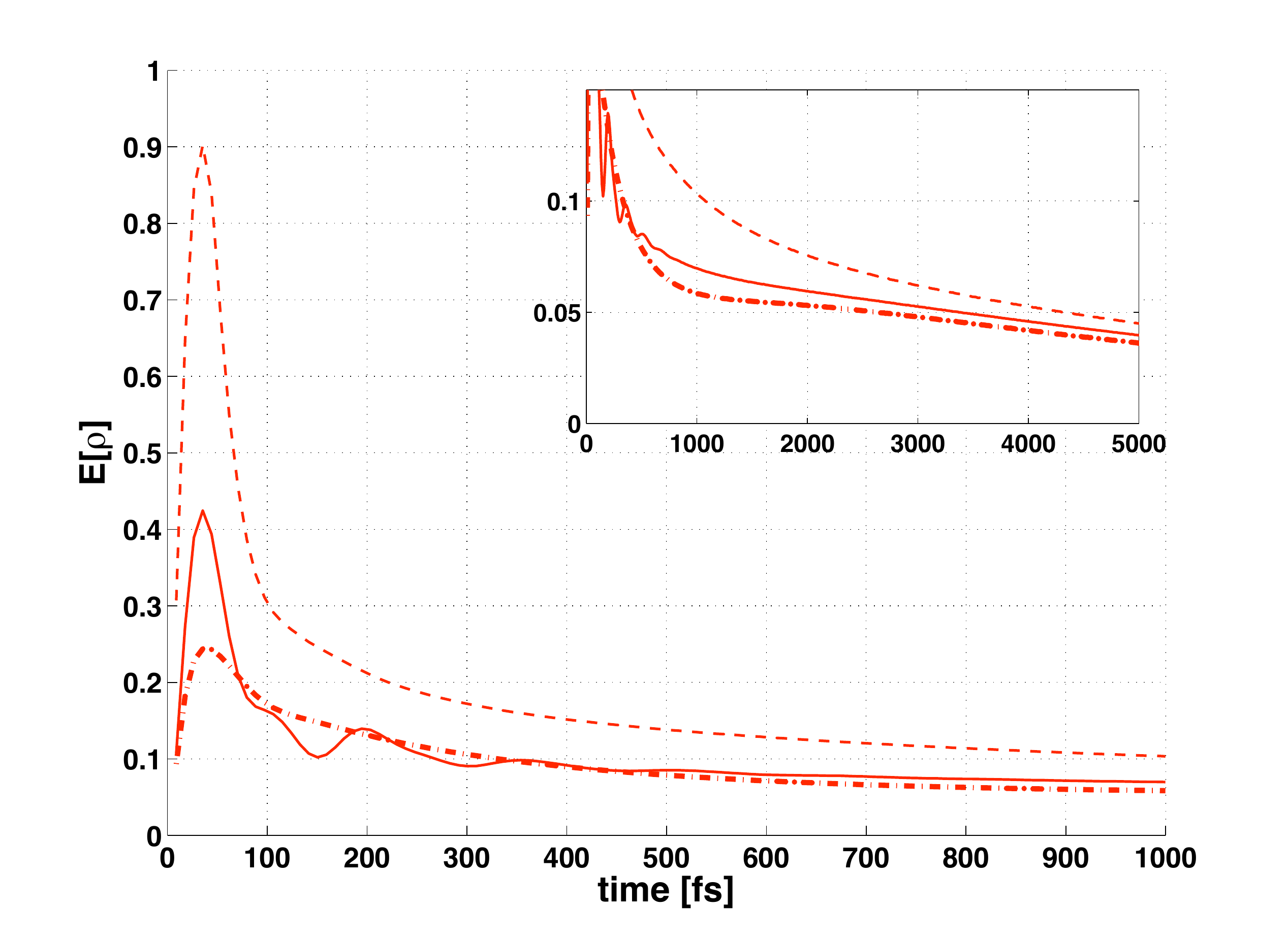}
\caption{\label{fig:redfield_comp} 
Comparison of time dependence of predicted global entanglement in the FMO complex for three different simulation models of the excitation dynamics. Solid curve: evolution of the entanglement measure $E[\rho]$ when the dynamics is generated with the I-F model of Ref. \cite{Ish.Fle-2009}, as in the figures in the main text. Dashed curve: evolution of $E[\rho]$ when the dynamics is generated by the full Redfield equation. Dash-dotted curve:  evolution of $E[\rho]$  with dynamics generated by the Redfield equation with the secular approximation. Inset shows the long time behavior of the same quantities.}
\end{figure}

Figure \ref{fig:redfield_comp} shows the time evolution of the global entanglement measure $E[\rho]=- \sum_{i=1}^N \rho_{ii}\ln \rho_{ii} - S(\rho)$ (where $S[\rho]$ is the von Neumann entropy of $\rho$), at T=300K, when the initial state is $\ket{1}$, for three models of FMO dynamics: the Ishizaki and Fleming (I-F) model \cite{Ish.Fle-2009}, the Redfield model with the secular approximation, and the full Redfield equation. We see that on the whole, the full Redfield model overestimates, while the Redfield model with the secular approximation underestimates entanglement in the system. Notice that neither of these approximate Redfield models capture the oscillations in global entanglement that our theory predicts with the simulations according to the I-F equations. This is due to the fact that, unlike the I-F model, neither Redfield model realistically accounts for the phonon relaxation dynamics that occur at timescales comparable to excitation dynamics \cite{Ish.Fle-2009a}. 

\section{Fenna-Matthews-Olson complex electronic Hamiltonian}
The FMO protein is a well characterized light harvesting complex. We use the \textit{Chlorobium tepidum} site energies and coupling strengths from table 4 (the trimer column) and table 1 (column (4)) of Ref. \cite{Ado.Ren-2006} to form a Hamiltonian that describes the closed-system dynamics of an FMO monomer excitation. Explicitly, the Hamiltonian matrix takes the form:
\beq
H_{\textrm{el}} = \begin{pmatrix} 
 200 & -87.7 & 5.5 & -5.9 & 6.7 & -13.7 & -9.9 \\
 -87.7 & 320 & 30.8 & 8.2 & 0.7 & 11.8 & 4.3 \\
 5.5 & 30.8 & 0 & -53.5 & -2.2 & -9.6 & 6.0 \\
 -5.9 & 8.2 & -53.5 & 110 & -70.7 & -17.0 & -63.3 \\
  6.7 & 0.7 & -2.2 & -70.7 & 270 & 81.1 & -1.3 \\
 -13.7 & 11.8 & -9.6 & -17.0 & 81.1 & 420 & 39.7 \\
 -9.9 & 4.3 & 6.0 & -63.3 & -1.3 & 39.7 & 230 
\end{pmatrix}
\eeq
in units of $\textrm{cm}^{-1}$ and with a constant offset of $12,210 \textrm{cm}^{-1}$ to set the lowest site energy to zero for convenience (this overall shift in energy does not affect the dynamics of the system).

\vspace{2cm}

\textbf{\textit{Acknowledgements}}
We are grateful to Yuan-Chung Cheng, Jahan Dawlaty, Vlatko Vedral and Martin Plenio for helpful conversations and comments. 
This material is based upon work supported by DARPA under Award No. N66001-09-1-2026. This work was supported by the Director, Office of Science, Office of Basic Energy Sciences, of the U.S. Department of Energy under Contract No. DE-AC02-05CH11231 and by the Chemical Sciences, Geosciences and Biosciences Division, Office of Basic Energy Sciences, U.S. Department of Energy under contract DE-AC03-76SF000098. A.I. appreciates the support of the Japan Society for the Promotion of Science (JSPS) Postdoctoral Fellowship for Research Abroad.

\end{document}